\documentclass[prd,showpacs,preprintnumbers,amsmath,amssymb]{revtex4}
\usepackage{amsmath,amssymb,amsthm}
\usepackage{color}
\usepackage{epsfig}
\usepackage{multirow}
\usepackage{graphicx}
\usepackage{subfigure}
\usepackage{rotating}

\begin{document}

%<<<<<<<<<<<<< AUTHOR >>>>>>>>>>>>>>>%  
\title{Strong gravitational lensing of a 4-dimensional Einstein-Gauss-Bonnet black hole in homogeneous plasma}

\author{Xing-Hua Jin$^{1}$ } \email{jinxh@sbs.edu.cn}

\author{Yuan-Xing Gao$^{2}$} 

\author{Dao-Jun Liu$^{2}$} \email{djliu@shnu.edu.cn}

\affiliation{$^{1}$ Department of Mathematics, Shanghai Business School, Shanghai 200235, China}
\affiliation{$^{2}$ Center for Astrophysics and Department of Physics, Shanghai Normal University, Shanghai 200234, China}
%<<<<<<<<<<<<< AUTHOR >>>>>>>>>>>>>>>%  

\begin{abstract}
We investigate the strong gravitational lensing of spherically symmetric black holes in the novel Einstein-Gauss-Bonnet(EGB) gravity surrounded by unmagnetised plasma medium.
The deflection angle in the strong deflection limit in EGB spacetime with homogeneous plasma is derived.
We find that both the coupling constant $\alpha$ in the novel EGB gravity and the presence of plasma can affect the radius of photon sphere, strong field limit coefficient and other lensing observables significantly.
While plasma has little effect on the angular image separation and the relative magnifications as $\alpha/M^2\to -8$ 
and $\alpha/M^2\to 1$, respectively.
\end{abstract}

\pacs{04.70.Dy, 95.30.Sf, 98.62.Sb, 94.20.ws} 

\keywords{strong gravitational lensing; EGB spacetime; plasma; black hole.}

\maketitle

\section{Introduction}

Gavitational lensing is the phenomenon of deflection of light rays in a
gravitational field, which has been successfully employed to explain the astronomical observations in the weak field approximation \cite{Schneider_Ehlers_Falco_1992,Petters_Levine_Wambsganss_2001,Schneider_Kochanek_Wambsganss_2006, Bartelmann:2010fz} when deflection angle is small. 
When the light rays approach towards the photon sphere of black hole where the gravitational field is extremely strong, the deflection angle becomes so large that the weak field method is no longer valid. 
It was first noticed by Darwin \cite{Darwin} in 1959 that the light rays passing very close to a black hole would make complete one or more loops around it before falling into the event horizon, hence an infinite series of exotic images were produced.
Later, strong gravitational lensing was regained wide attention \cite{Atkinson, Luminet:1979nyg, Ohanian, Nemiroff:1993he}. 
The  exact lensing equation with arbitary large value of deflection angle is obtained in 2000 \cite{Frittelli:1999yf, Virbhadra:1999nm}. In 2001 Bozza et al. \cite{Bozza:2001xd} developed a reliable and analytical method to obtain the deflection angle of Schwarzschild black hole in strong field region and they found the logarithmic divergence of the deflection angle in strong field limit. Later Bozza \cite{Bozza:2002zj} extended the conclusion to a general asymptotically flat, static, and spherically symmetric spacetime.
With the help of strong gravitational lensing it is possible to compare alternative theories of gravity \cite{Claudel:2000yi, Hasse:2001by, Iyer:2006cn, Virbhadra:2007kw, Bozza:2008ev, Bozza:2009yw, Ghosh:2010uw, Wei:2011nj, Chen:2009eu} and pick up information from different compact objects \cite{Tsukamoto:2016qro, Bozza:2002af, Vazquez:2003zm, Bozza:2005tg, Bozza:2006nm, Chen:2010yx, Chen:2011ef, Cunha:2015yba, Cavalcanti:2016mbe, Gyulchev:2008ff, Sahu:2012er, Sahu:2013uya, Kuhfittig:2013hva, Nandi:2006ds, Tsukamoto:2012xs, Tsukamoto:2016jzh}.
Last year, the first image of the supermassive black hole M87$^{*}$ at the center of the galaxy M87 has been captured by the Event Horizon Telescope (EHT)\cite{Akiyama:2019cqa, Akiyama:2019fyp, Akiyama:2019eap}, which provides us the deeper understanding of the strong gravitational physics.

One of the simplest natural extension of Einstein's gravity by higher curvature correction is the Einstein-Gauss-Bonnet (EGB) gravity, the action of which in $D$-dimensional spacetime is given by
\begin{equation}
S
= \frac{1}{16\pi}\int d^{D}x \sqrt{-g}
\left[\frac{M_{\rm P}^2}{2}R+\alpha\mathcal{G}\right],
\end{equation}
where $\alpha$ is the coupling constant of the Gauss-Bonnet (GB) term
\begin{equation}
\mathcal{G}= {R^{\mu\nu}}_{\rho\sigma} {R^{\rho\sigma}}_{\mu\nu}-
4 {R^\mu}_\nu {R^\nu}_\mu + R^2 = 
6 {R^{\mu\nu}}_{[\mu\nu} {R^{\rho\sigma}}_{\rho\sigma]},
\end{equation}
with $R_{\mu\nu\rho\sigma}$ the Riemann tensor, $R_{\mu\nu}$ the Ricci tensor and $R$ the Ricci scalar.
In the  $4$-dimensional spacetime, GB term is a total derivative \cite{Lanczos:1938sf}, so it has no contribution to the gravitational dynamics.
However, the role of the GB term in $4$-dimensional gravity, in particular, holographic implications to the addition of it to the gravity action was studied in Ref. \cite{Miskovic:2009bm}.
Notice that standard thermodynamics for AdS black holes is recovered in this way.
Recently, Glavan and Lin \cite{Glavan:2019inb} reformulate the $D$-dimensional EGB gravity by rescaling the coupling $\alpha\rightarrow\alpha/(D-4)$. They obtain a novel $4$-dimensional EGB gravity theory in the limit $D\rightarrow4$, where the GB term can give the nontrivial contribution of gravitational dynamics. They also have shown that it can bypass the Lovelock's theorem \cite{Lovelock:1971yv,Lovelock:1972vz} and prevent Ostrogradsky instability \cite{Woodard:2015zca}.
This idea of regularization can be traced back to Refs \cite{Tomozawa:2011gp, Cognola:2013fva}, which gives the quantum corrections of Einstein's gravity.
In addition, a novel static spherically symmetric black hole solution  was obtained within this theory.
Note that the black hole solution was found earlier in the gravity theories with conformal anomaly \cite{Cai:2009ua} and quantum corrections \cite{Tomozawa:2011gp, Cognola:2013fva}, and recently in regularized Lovelock gravity \cite{Casalino:2020kbt}, respectively.

The novel $4$-dimensional EGB black holes are free from singularity problem. Their photon sphere and shadow, as well as the innermost stable circular orbit (ISCO) of a spinless test particle \cite{Guo:2020zmf} and spinning test particle \cite{Zhang:2020qew} around them, have been calculated. 
Quasinormal modes of bosonic fields \cite{Konoplya:2020bxa} and fermionic fields \cite{Churilova:2020aca} of these black holes have been investigated, and it is found that for the bosonic fields the damping rate is more sensitive than the real part of quasinormal modes by changing of the GB coupling constant $\alpha$,
while for the fermionic fields the damping rate usually decreases and the real part of the quasinormal modes increases with the increase of $\alpha$. Konoplya and Zhidenko discussed the stability \cite{Konoplya:2020juj} of spherically symmetric black holes in the novel EGB gravity. Moreover, other topics in this new theory including the charged black holes in AdS spaces \cite{Fernandes:2020rpa}, the shadow of dS black holes \cite{Roy:2020dyy}, the bending of light in dS black holes \cite{Heydari-Fard:2020sib}, the rotating black holes \cite{Wei:2020ght,Kumar:2020owy}, radiating black holes \cite{Ghosh:2020vpc},  the structure of relativistic stars \cite{Doneva:2020ped}, the thermodynamics of the black holes \cite{Hegde:2020xlv, Singh:2020xju, Zhang:2020qam, HosseiniMansoori:2020yfj} and the accretion disk around the black hole \cite{Liu:2020vkh} have also been studied.
However, several problems, such as completeness, about the regularization procedure have been put foward in Refs \cite{Ai:2020peo, Gurses:2020ofy, Shu:2020cjw, Hennigar:2020lsl, Mahapatra:2020rds, Tian:2020nzb,Ge:2020tid}, in the meantime some prescriptions have been suggested \cite{Casalino:2020kbt, Hennigar:2020lsl, Lu:2020iav,  Kobayashi:2020wqy}.	L\"u and Pang \cite{Lu:2020iav} proposed a more rigorous way to regularize the EGB gravity by compactifying the $D$ dimensional EGB gravity on the $(D-4)$ dimensional maximally symmetric space and redefining the coupling constant as $\alpha/(D-4)$. 
In accordance with the results of Ref. \cite{Kobayashi:2020wqy}, a special scalar-tensor theory that belongs to the family of Horndeski gravity is obtained by this method.
Ref. \cite{Hennigar:2020lsl} extends the method for obtaining the $D\rightarrow2$ limit of general relativity \cite{Mann:1992ar} to the $D\rightarrow4$ limit of EGB gravity.
Anyhow, in these regularised theories \cite{Casalino:2020kbt, Hennigar:2020lsl, Lu:2020iav} the spherically symmetric 4D black hole solution obtained in Refs \cite{Glavan:2019inb, Cognola:2013fva}  is still valid.

On the other hand, it is believed that there exists plasma fluid surrounding black holes and other compact objects. When the light moves towards the compact objects through the plasma, the trajectory of light is different from the vacuum case. The theory of the light propagation in a curved spacetime in the presence of an isotropic dispersive medium was considered in the classical book of Synge \cite{Synge}. Synge used the general relativistic Hamiltonian approach to deal with the geometrical optics in a dispersive medium.
Furthermore, the influence of a spherically symmetric and time-independent plasma on the light defection in Schwarzschild spacetime and Kerr spacetime was discussed in the book of Perlick \cite{Perlick1}.
The effect of plasma on the shadows of black holes and wormholes has been investigated in \cite{Bisnovatyi-Kogan:2017kii, Abdujabbarov:2015pqp, Perlick:2017fio, Abdujabbarov:2016efm, Huang:2018rfn}. 
Gravitational lensing by the compact object in homogeneous and inhomogeneous plasma was considered in \cite{BisnovatyiKogan:2008yg, BisnovatyiKogan:2010ar, Morozova, Er:2013efa, Atamurotov:2015nra, Rogers, Perlick:2015vta, Tsupko:2013cqa}.

In this work, we shall study the strong gravitational lensing by this novel $4$-dimensional EGB black hole in an unmagnetized homogeneous plasma medium. The rest of the paper is organized as follows.
In Sec.~II, we study the photon sphere radius and the critical value of impact parameter of this novel black hole in the presence of plasma and derive the expression for the deflection angle of light in Sec.~III.
In Sec.~IV, we investigate the effects of  plasma on the deflection angle, the coefficients and the observable quantities for gravitational lensing in the strong field limit. Finally, We end the paper with a summary in Sec.~V.
Throughout this paper we use the units in which $G=c=1$.

\section{Photon sphere of an Einstein-Gauss-Bonnet black hole in the presence of plasma}

Let us start from the line element of the EGB black hole spacetime \cite{Glavan:2019inb}, which is given by
\begin{eqnarray}\label{metric}
ds^{2}=-A(r)dt^{2}+B(r)dr^{2}+C(r)(d\theta^{2}+\sin^{2}\theta d\varphi^{2}),
\end{eqnarray}
where the functions $A(r)$, $B(r)$ and $C(r)$ have respectively the following form,
\begin{eqnarray}
&&A(r)= 1 + \frac{r^2}{2\alpha}
\Biggl( 1- \sqrt{1+\frac{8\alpha M}{r^3}}\Biggr),\label{A}\\
&&B(r)=\Biggl[1 + \frac{r^2}{2 \alpha}
\Biggl( 1- \sqrt{1+\frac{8\alpha M}{r^3}}\Biggr)\Biggr]^{-1},\label{B}\\
&&C(r)=r^2\label{C}.
\end{eqnarray}
It has been shown that the metric is asymptotic flat by the expansion at large $r$.
Here $M$ is the mass of the EGB black hole and the GB coupling constant $\alpha$  is constrained in the range $-8\le {\alpha}/{M^2}\le 1$ \cite{Guo:2020zmf}. For the case $0<{\alpha}/{M^2}\le 1$, there are two horizons 
\begin{equation}
r_{\pm}=M\pm \sqrt{M^2-\alpha}.
\end{equation}
While for the case $-8\le{\alpha}/{M^2}< 0$, there is only one horizon $r_{+}$,
where the singular short radial distances $r<\sqrt[3]{-8\alpha M}$ are concealed inside this outer horizon. We will take the region $-8\le {\alpha}/{M^2}\le 1$ for the coupling constant in this paper.
 
We assume that the spacetime is filled with a spherically symmetric distribution of plasma with electron plasma frequency
\begin{equation}
\omega_p(r)^2 = \frac{4\pi e^2}{m} N(r),
\end{equation}
where $e$ and $m$ are the charge of the electron and the mass of the electron respectively. The number density of the electrons $N(r)$ is the function of the radius coordinate only.
The relation between the refraction index $n$ and the photon frequency $\omega$ is given as
\begin{equation}
n ^2 = 1 - \frac{\omega_p^2(r)}{\omega^2}.
\end{equation}
It is found that when $\omega > \omega_p$, the photon can propagate through the plasma. On the other hand, when $\omega < \omega_p$, the photon motion is forbidden \cite{BisnovatyiKogan:2010ar, Rogers}.
Note that one has $n=1$ in the vacuum case. 

We start to calculate the strong gravitational lensing of the EGB black hole surrounded by plasma. The trajectories of photons in a curved space-time with plasma mediums, were obtained by Synge \cite{Synge}. The Hamiltonian for
the light rays around the black hole surrounded by plasma has the
following form \cite{Kulsrud:1991jt}
\begin{equation}\label{H}
H(x,p) = \frac{1}{2} \left[ g^{\mu\nu} p_{\mu} p_{\nu} +\omega_p^2(r)
\right] = 0,
\end{equation}
where $p_{\mu}$ is the four-momentum of the photon and 
$g^{\mu\nu}$ is the contravariant metric tensor.
Substituing (\ref{metric}) into (\ref{H}), we get the equation
\begin{equation}\label{H1}
0=-\frac{p_t^2}{A(r)}+\frac{p_r^2}{B(r)}+\frac{p_{\varphi}^2}{C(r)}+\omega_p^2(r).
\end{equation}
Using the Hamiltonian (\ref{H}) for the photon around
the EGB black hole, the paths of light rays are then described in terms of the
affine parameter $\lambda$ by
\begin{equation}\label{dxp} 
\frac{dx^{\mu}}{d \lambda} = \frac{\partial H}{\partial
	p_{\mu}},~ \frac{dp_{\mu}}{d \lambda} = - \frac{\partial
	H}{\partial x^{\mu}}.
\end{equation}
Because of the spherical symmetry, we can confine the photon orbits in the equatorial plane by taking $\theta=\pi/2$ without the loss of generality.
The coordinates $t$ and $\varphi$ are cyclic, leading two costants of motions which are the energy $E$ and the angular
momentum $L$ of the photon 
\begin{equation}
E=-p_t=\omega_{\infty},~ L=p_\varphi,
\end{equation}
where $\omega_{\infty}$ is the photon frequency at infinity. From Eqs. (\ref{metric}) and (\ref{dxp}), the expression for 
${dr}/{d\lambda}, ~{d\varphi}/{d\lambda}$ is obtained in terms of $p_r$ and $p_\varphi$ 
\begin{eqnarray}
\label{dr}&&\frac{dr}{d\lambda} =  \frac{\partial H}{\partial p_r}= 
\frac{p_r}{B(r)},\\
\label{dphi}&&\frac{d\varphi}{d\lambda} =  \frac{\partial H}{\partial p_{\varphi}} =  
\frac{p_{\varphi}}{C(r)}.
\end{eqnarray}
Using Eqs. (\ref{H1}), (\ref{dr}) and (\ref{dphi}), we obtain the equation of trajectory for a photon which is similar to the formalism in Ref. \cite{Tsukamoto:2016qro}
\begin{equation}\label{drphi}
\left(\frac{dr}{d\varphi}\right)^{2}=\frac{R_p(r)C(r)}{B(r)}
\end{equation}
where
\begin{eqnarray}
&&R_p=\frac{E^2}{L^2}\frac{C(r)}{A(r)}W(r)-1,\label{Rp}\\
&&W(r)=1 -\frac{\omega _p (r) ^2}{E^2}A(r).\label{W}
\end{eqnarray}
In the case $\omega_p(r)=0$ or equivalently, $W(r)=1$, Eq. (\ref{drphi}) gives the motion of light ray in vacuum.

We are interested in a photon with a given enegy $E$ that comes in from infinity, reaches a closest distance $r=r_{0}$, and goes out to infinity.
As $r_0$ corresponds to the turning point of the path, $dr/d\varphi$ vanishes and $R_p(r_0)=0$. Hereafter subscript $0$ indicates the quantity at the closest distance $r=r_{0}$. 
For a light ray initially in the asymptotically flat spacetime, the impact parameter can be represented as 
\begin{equation}\label{impact}
b(r_{0})=\frac{L}{E}=\sqrt{\frac{C_{0}W_{0}}{A_{0}}}. 
\end{equation}
With the help of Eq. (\ref{impact}), $R_p(r)$ can be rewritten as
\begin{equation}\label{R2}
R_p(r)= \frac{A_{0}CW}{AC_{0}W_{0}}-1.
\end{equation}
To find the radius of photon sphere, which is the unstable circular photon orbit of static, spherically symmetric compact objects, one can introduce a function $h(r)$ given by Perlick \cite{Perlick:2015vta}
\begin{equation}\label{h}
h(r)^2=\frac{C(r)}{A(r)}W(r)=\frac{C(r)}{A(r)}\left[1 -\frac{\omega _p (r) ^2}{E^2}A(r)\right]. 
\end{equation}
The photon sphere radius $r_m$ is the biggest real root of the equation
\begin{equation}\label{dh}
\frac{d}{dr}h(r)^2=0.
\end{equation}
From Eq. (\ref{dh}), we obtain
\begin{equation}\label{Dp}
\frac{C'}{C}+\frac{W'}{W}-\frac{A'}{A}=0,
\end{equation}
where prime denotes the differentiantion with respect to the radical coordinate $r$.

\begin{figure}
	\includegraphics[width=80mm,angle=0]{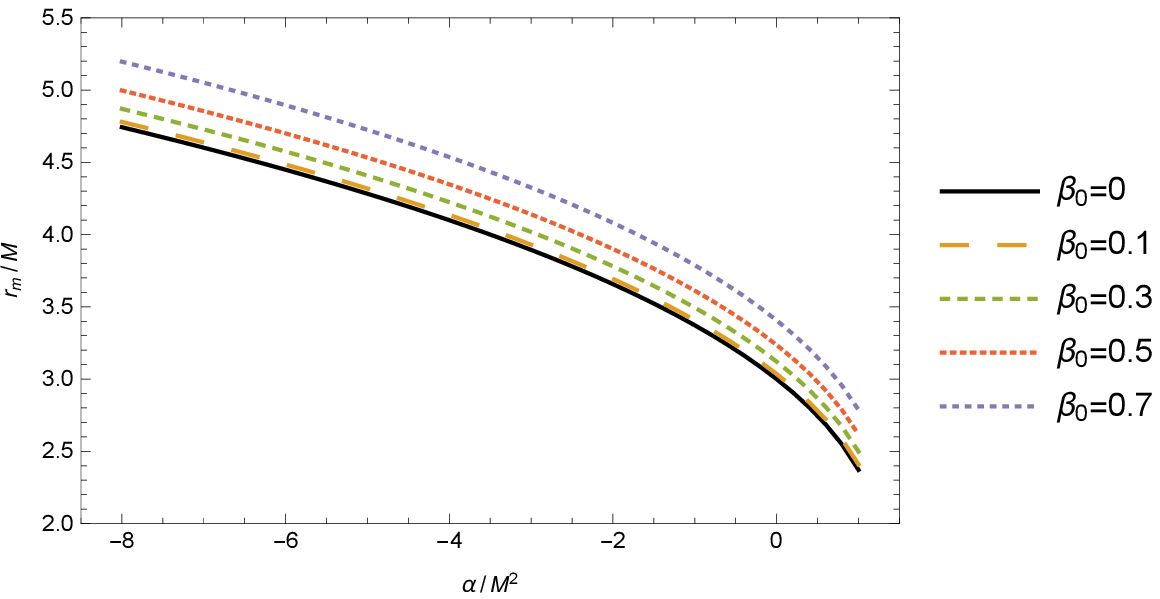}\,
	\includegraphics[width=80mm,angle=0]{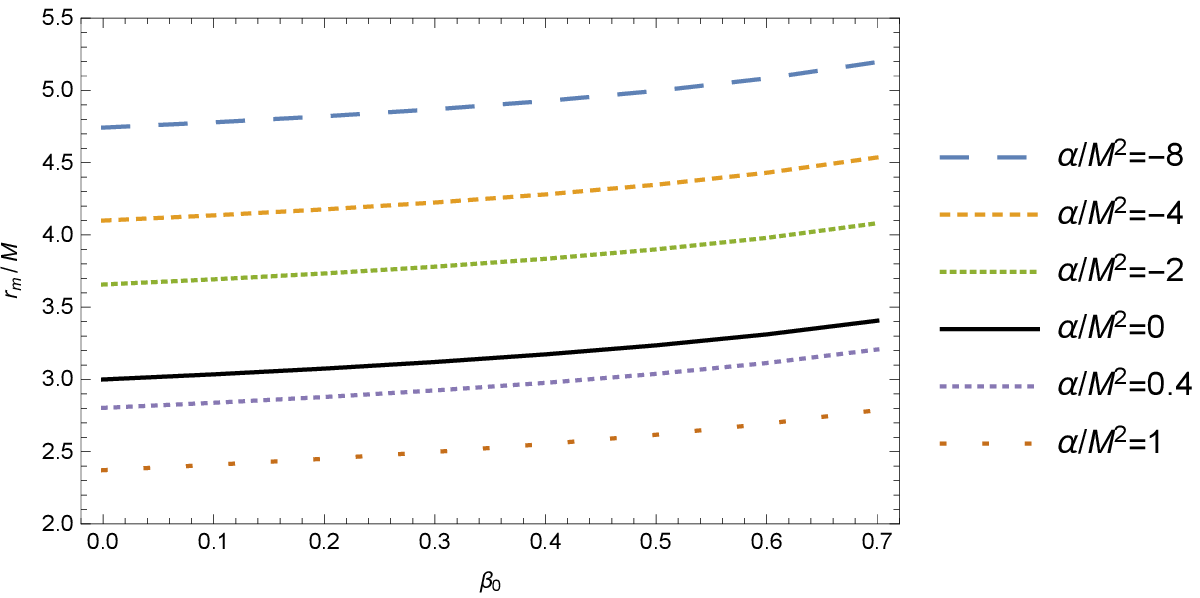}
\caption{Left panel: The plot of the radius of the photon sphere $r_m/M$ as a function of $\alpha/M^2$. The five curved lines are plotted when $\beta_0=0$, $\beta_0=0.1$, $\beta_0=0.3$, $\beta_0=0.5$ and $\beta_0=0.7$ respectively.
Right panel: The plot of the radius of the photon sphere $r_m/M$ as a function of $\beta_0$. The six curved lines are plotted when $\alpha/M^2=-8$, $\alpha/M^2=-4$, $\alpha/M^2=-2$, $\alpha/M^2=0$, $\alpha/M^2=0.4$ and $\alpha/M^2=1$ respectively.}
	\label{rm}
\end{figure}

Now we consider the EGB black hole surrounded by homogeous plasma, which has the following form
\begin{equation}\label{beta0}
\frac{\omega_p(r)}{E}=\beta_0,
\end{equation}
where $\beta_0$ is a positive dimensionless constant. Then we rewrite Eq. (\ref{Dp}) as
\begin{equation}\label{eqrm}
r\left[\beta _0 \left(2 \alpha ^2+r^4+2 \alpha  r^2+4 \alpha
Mr\right)-2 \alpha ^2 \right]\sqrt{\frac{8 \alpha M}{r^3}+1}=\beta _0 \left(16 \alpha^2M+r^5+2 \alpha  r^3+8 \alpha M
r^2\right)-6 \alpha ^2M.
\end{equation}
We can solve this equation numerically to get the radius of the photon sphere  which is plotted in Fig. \ref{rm}.
In the left panels of Fig. \ref{rm} we
show the function $r_m/M$ for $\beta_0=0$, $\beta_0=0.1$, $\beta_0=0.3$, $\beta_0=0.5$ and $\beta_0=0.7$ respectively, 
and we demonstrate that the radius of the photon sphere of the EGB black hole decreases with the increase of $\alpha/M^2$ for fixed $\beta_0$. 
In the right panels of Fig. \ref{rm} we
show the function $r_m/M$ for $\alpha/M^2=-8$, $\alpha/M^2=-4$, $\alpha/M^2=-2$, $\alpha/M^2=0$, $\alpha/M^2=0.4$ and $\alpha/M^2=1$ respectively, 
and we find that the radius of the photon sphere of the EGB black hole increases with the increase of $\beta_0$ for fixed $\alpha/M^2$.
It is clear that the presence of coupling constant $\alpha$ and the plasma parameter $\beta_0$, affects the photon sphere radius significantly.
 In the absence of $\beta_0$, from Eq. (\ref{eqrm}), the largest real root has a form
\begin{equation}
r_{m}=2 \sqrt{3}M \cos \left[\frac{1}{3} \cos ^{-1}\left(-\frac{4 \alpha }{3 \sqrt{3}M^2}\right)\right],
\end{equation}
which is the photon radius of the EGB black hole in vacuum \cite{Guo:2020zmf}.
On the other hand, in the case $\alpha= 0$, we can get the photon radius of Schwarzschild black hole with homogeneous plasma   
\begin{equation}
r_{m}=\frac{3-4 \beta_0 +\sqrt{9-8 \beta _0}}{2\left(1-\beta _0\right)}M,
\end{equation}
which has been obtained in Ref. \cite{Tsupko:2013cqa}.
   
\begin{figure}
\includegraphics[width=80mm,angle=0]{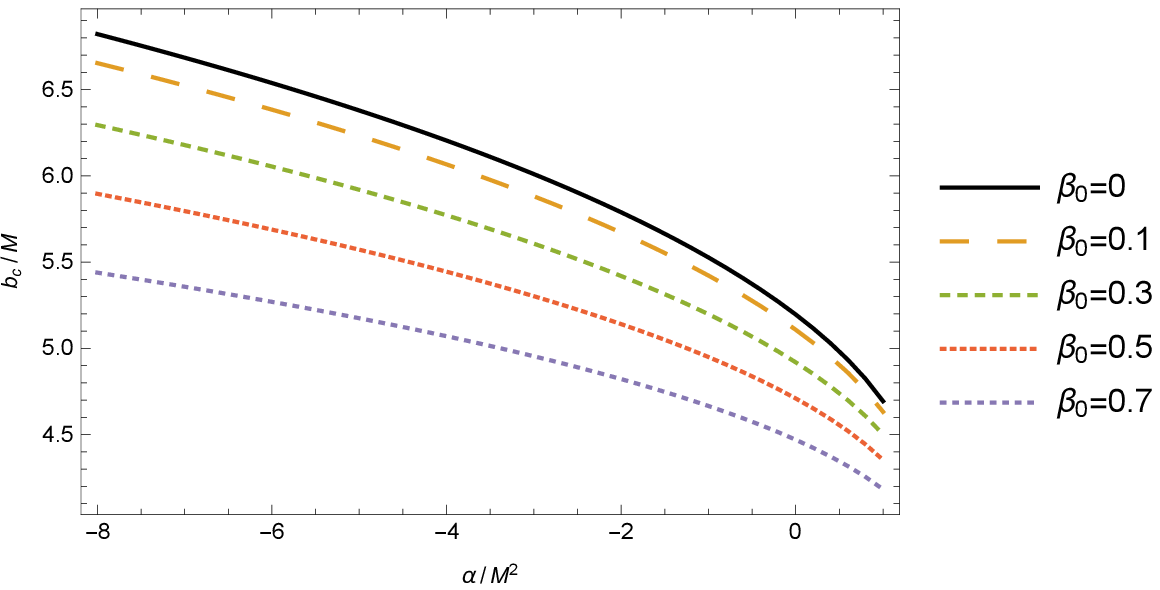}\,
\includegraphics[width=80mm,angle=0]{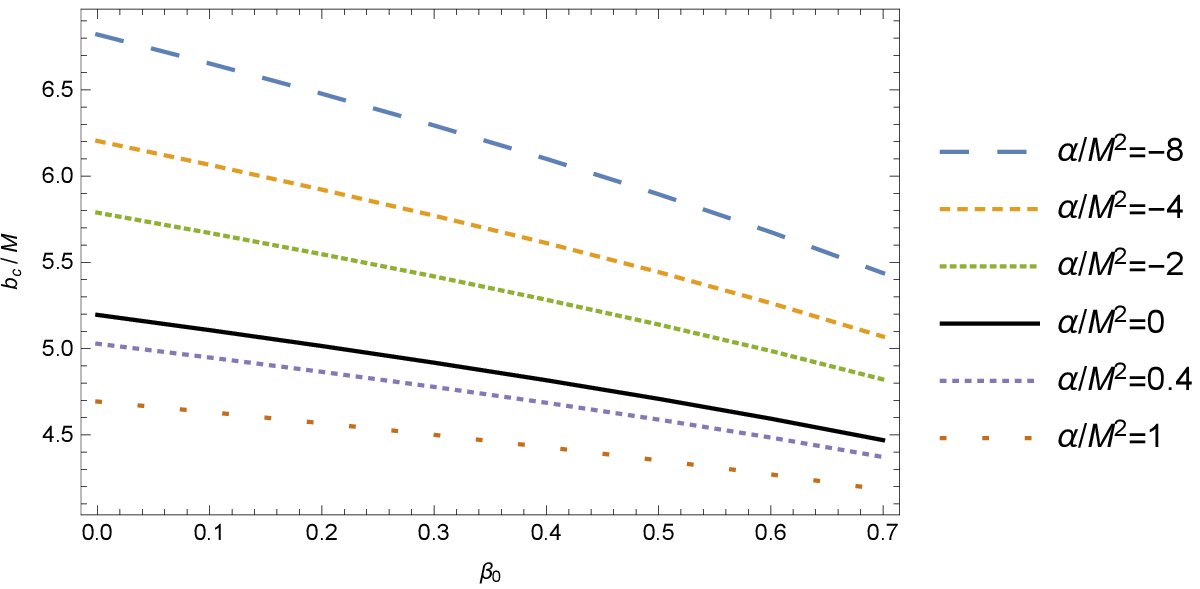}
\caption{Left panel: The plot of the critical impact parameter $b_c/M$ as a function of $\alpha/M^2$. The five curved lines are plotted when $\beta_0=0$, $\beta_0=0.1$, $\beta_0=0.3$, $\beta_0=0.5$ and $\beta_0=0.7$ respectively.
Right panel: The plot of the critical impact parameter $b_c/M$ as a function of $\beta_0$. The six curved lines are plotted when $\alpha/M^2=-8$, $\alpha/M^2=-4$, $\alpha/M^2=-2$, $\alpha/M^2=0$, $\alpha/M^2=0.4$ and $\alpha/M^2=1$ respectively.}
	\label{bc}
\end{figure}

We define the critical value of the impact parameter $b_{c}$ for the light ray as 
\begin{equation}\label{cip}
b_{c}
\equiv \lim_{r_{0}\rightarrow r_{m}} \sqrt{\frac{C_{0}W_{0}}{A_{0}}}.
\end{equation}
The strong deflection limit corresponds to the limit $r_0\rightarrow r_m$ or $b\rightarrow b_c$. From Eqs. (\ref{A}), (\ref{C}) and (\ref{W}), the critical impact parameter is given by
\begin{eqnarray}
b_{c}(r_{m})=\sqrt{\frac{\beta _0 \left[r_m^4 \left(\sqrt{\frac{8 \alpha  M}{r_m^3}+1}-1\right)-2 \alpha r_m^2\right]+2 \alpha  r_m^2}{2 \alpha +r_m^2 \left(1-\sqrt{\frac{8 \alpha	M}{r_m^3}+1}\right)}}.
\end{eqnarray}
The dependence of the critical impact parameter from  the coupling constant and the plasma parameters is shown in Fig. \ref{bc}.
The left panels of Fig. \ref{bc} presents the function $b_c/M$ for $\beta_0=0$, $\beta_0=0.1$, $\beta_0=0.3$, $\beta_0=0.5$ and $\beta_0=0.7$ respectively, 
and it shows that the critical impact parameter of the EGB black hole decreases with the increase of $\alpha/M^2$ for fixed $\beta_0$. 
The right panels of Fig. \ref{bc} presents the function $b_c/M$ for $\alpha/M^2=-8$, $\alpha/M^2=-4$, $\alpha/M^2=-2$, $\alpha/M^2=0$, $\alpha/M^2=0.4$ and $\alpha/M^2=1$ respectively, and it shows that the critical impact parameter of the EGB black hole decreases with the increase of $\beta_0$ for fixed $\alpha/M^2$.
We found that both the coupling constant and the presence of plasma have remarkable influences on the critical impact parameter.

\section{Strong gravitational lensing of EGB black hole in homogeneous plasma}

In this section, we will calculate the deflection angle of a light ray in the strong deflction limit in the EGB black hole spacetime with plasma medium.
From Eq. (\ref{drphi}), the deflection angle $\hat{\alpha}_p(r_{0})$ for the photon coming from infinite to the EGB black hole in homogeneous plasma is given by
\begin{equation}\label{alpha}
\hat{\alpha}_p(r_{0})=I_p(r_{0})-\pi,
\end{equation}
where $I_p(r_{0})$ is defined as
\begin{equation}\label{angle2}
I_p(r_{0})\equiv 2\int^{\infty}_{r_{0}}\frac{1}{\sqrt{\frac{R_p(r)C(r)}{B(r)}}}dr.
\end{equation}
It is found that the deflection angle increases when the closest distance $r_0$ decreases, and for a special point, the deflection angle will arrive at $2\pi$ which means the photon winds a complete loop around the black hole. Furthermore, when $r_{0}$ approach the radius of the photon sphere $r_m$ the  deflection angle will diverge \cite{Virbhadra:1999nm}.
To discuss the divergence, following Ref.\cite{Chen:2009eu}, we  introduce a new variable $z$ 
\begin{equation}\label{z1}
z\equiv 1-\frac{r_{0}}{r}.
\end{equation}
Using Eqs. (\ref{A})-(\ref{C}), (\ref{Rp}) and (\ref{W}), we can rewrite $I_p(r_{0})$ as
\begin{equation}
I_p(r_{0})=\int^{1}_{0}f_p(z,r_{0})dz
=\int^{1}_{0}\frac{2r_0}{\sqrt{G_p(z,r_{0})}}dz,
\end{equation}
where the function $G_p(z,r_{0})$ in the EGB spacetime is given by
\begin{eqnarray}
G_p(z,r_{0})=&&\frac{R_p(z,r_0)C(z,r_0)}{B(z,r_0)}(1-z)^4=r_0^2\left((1-z)^2-\frac{r_0^2 \left(\sqrt{1-\frac{8 \alpha  M (z-1)^3}{r_0^3}}-1\right)}{2 \alpha}\right)
\\\nonumber
&&\left(\frac{\left(r_0^2
			\left(\sqrt{\frac{8 \alpha  M}{r_0^3}+1}-1\right)-2 \alpha \right) \left(-\beta _0 \left(r_0^2 \left(\sqrt{\frac{r_0^3-8 \alpha  M
					(z-1)^3}{r_0^3}}-1\right)-2 \alpha  (z-1)^2\right)-2 \alpha  (z-1)^2\right)}{(z-1)^2 \left(2 \alpha +\beta _0 \left(r_0^2 \left(\sqrt{\frac{8 \alpha
					M}{r_0^3}+1}-1\right)-2 \alpha \right)\right) \left(2 \alpha  (z-1)^2-r_0^2 \left(\sqrt{\frac{r_0^3-8 \alpha  M
					(z-1)^3}{r_0^3}}-1\right)\right)}-1\right).
\end{eqnarray}
We can expand the above expression into a power series of z in the following form
\begin{equation}
G_p(z,r_{0})=\sum_{n=1}^{\infty}c_n(r_0)z^n,
\end{equation}
where $c_{1}(r_{0})$ and $c_{2}(r_{0})$ are given by
\begin{eqnarray}
c_{1}(r_{0})=&&\frac{\beta _0 \left(-4 \alpha ^2 r_0^2 \sqrt{\frac{8 \alpha  M}{r_0^3}+1}+32 \alpha ^2 M r_0+\alpha  r_0^4 \left(4-4 \sqrt{\frac{8 \alpha 
			M}{r_0^3}+1}\right)-8 \alpha  M r_0^3 \left(\sqrt{\frac{8 \alpha  M}{r_0^3}+1}-2\right)\right)}{\alpha  \sqrt{\frac{8 \alpha  M}{r_0^3}+1} \left(2
	\alpha +\beta _0 \left(r_0^2 \left(\sqrt{\frac{8 \alpha  M}{r_0^3}+1}-1\right)-2 \alpha \right)\right)}
\\\nonumber
&&+\frac{\beta _0 r_0^6 \left(2-2 \sqrt{\frac{8 \alpha M}{r_0^3}+1}\right)+4 \alpha ^2 r_0 \left(r_0 \sqrt{\frac{8 \alpha  M}{r_0^3}+1}-3 M\right)}{\alpha  \sqrt{\frac{8 \alpha  M}{r_0^3}+1} \left(2 \alpha +\beta _0 \left(r_0^2 \left(\sqrt{\frac{8 \alpha	M}{r_0^3}+1}-1\right)-2 \alpha \right)\right)},
\end{eqnarray}
and 
\begin{eqnarray}\label{c2}
c_{2}(r_{0})=&&\frac{2 \alpha ^2 {r_0}^2 \left(96 \alpha ^2 M^3 {r_0}+{r_0}^8 \left(-\sqrt{\frac{8 \alpha  M}{{r_0}^3}+1}\right)+{r_0}^8+2 \alpha 
	{r_0}^6\right)}{{\alpha  \left(8 \alpha  M+{r_0}^3\right)^2 \left(2 \alpha +{r_0}^2 \left(1-\sqrt{\frac{8 \alpha  M}{{r_0}^3}+1}\right)\right) \left(\beta
		_0 \left(2 \alpha +{r_0}^2 \left(1-\sqrt{\frac{8\alpha M}{{r_0}^3}+1}\right)\right)-2 \alpha \right)}}
\\\nonumber
&&+\frac{2 \alpha ^2 M^2 {r_0}^2 \left(128 \alpha ^3+\alpha  {r_0}^4 \left(60-12 \sqrt{\frac{8 \alpha  M}{{r_0}^3}+1}\right)+\alpha ^2 {r_0}^2
	\left(64-88 \sqrt{\frac{8 \alpha  M}{{r_0}^3}+1}\right)\right)}{{\alpha  \left(8 \alpha  M+{r_0}^3\right)^2 \left(2 \alpha +{r_0}^2 \left(1-\sqrt{\frac{8 \alpha  M}{{r_0}^3}+1}\right)\right) \left(\beta
		_0 \left(2 \alpha +{r_0}^2 \left(1-\sqrt{\frac{8 \alpha  M}{{r_0}^3}+1}\right)\right)-2 \alpha \right)}}
\\\nonumber
&&+\frac{2 \alpha ^2 M {r_0}^2 \left({r_0}^7 \left(6-6 \sqrt{\frac{8 \alpha  M}{{r_0}^3}+1}\right)+\alpha  {r_0}^5 \left(16-28 \sqrt{\frac{8
			\alpha  M}{{r_0}^3}+1}\right)+32 \alpha ^2 {r_0}^3\right)}{{\alpha  \left(8 \alpha  M+{r_0}^3\right)^2 \left(2 \alpha +{r_0}^2 \left(1-\sqrt{\frac{8 \alpha  M}{{r_0}^3}+1}\right)\right) \left(\beta
		_0 \left(2 \alpha +{r_0}^2 \left(1-\sqrt{\frac{8 \alpha  M}{{r_0}^3}+1}\right)\right)-2 \alpha \right)}}
\\\nonumber&&+\frac{-2 \beta _0 {r_0}^2 \left(8 \alpha  M+{r_0}^3\right) \left(2 \alpha ^2 {r_0}^3 \left(\alpha -24 M^2\right)+32 \alpha ^3 M^2 {r_0}+16
	\alpha ^4 M+\alpha ^2 {r_0}^5 \left(\sqrt{\frac{8 \alpha  M}{{r_0}^3}+1}-1\right)\right)}{\alpha  \left(8 \alpha  M+{r_0}^3\right)^2 \left(2 \alpha +{r_0}^2 \left(1-\sqrt{\frac{8 \alpha  M}{{r_0}^3}+1}\right)\right) \left(\beta
	_0 \left(2 \alpha +{r_0}^2 \left(1-\sqrt{\frac{8 \alpha  M}{{r_0}^3}+1}\right)\right)-2 \alpha \right)}
\\\nonumber&&+\frac{-2 \beta _0 {r_0}^2 \left(8 \alpha  M+{r_0}^3\right) \left(6 \alpha  M {r_0}^6 \left(3 \sqrt{\frac{8 \alpha  M}{{r_0}^3}+1}-5\right)-8
	\alpha ^3 M {r_0}^2 \left(2 \sqrt{\frac{8 \alpha  M}{{r_0}^3}+1}+1\right)\right)}{\alpha  \left(8 \alpha  M+{r_0}^3\right)^2 \left(2 \alpha +{r_0}^2 \left(1-\sqrt{\frac{8 \alpha  M}{{r_0}^3}+1}\right)\right) \left(\beta
	_0 \left(2 \alpha +{r_0}^2 \left(1-\sqrt{\frac{8 \alpha  M}{{r_0}^3}+1}\right)\right)-2 \alpha \right)}
\\\nonumber&&+\frac{-2 \beta _0 {r_0}^2 \left(8 \alpha  M+{r_0}^3\right) \left({r_0}^9 \left(3 \sqrt{\frac{8 \alpha  M}{{r_0}^3}+1}-3\right)+\alpha 
	{r_0}^7 \left(5 \sqrt{\frac{8 \alpha  M}{{r_0}^3}+1}-5\right)+4 \alpha ^2 M {r_0}^4 \left(4 \sqrt{\frac{8 \alpha 
			M}{{r_0}^3}+1}-9\right)\right)}{\alpha  \left(8 \alpha  M+{r_0}^3\right)^2 \left(2 \alpha +{r_0}^2 \left(1-\sqrt{\frac{8 \alpha  M}{{r_0}^3}+1}\right)\right) \left(\beta
	_0 \left(2 \alpha +{r_0}^2 \left(1-\sqrt{\frac{8 \alpha  M}{{r_0}^3}+1}\right)\right)-2 \alpha \right)}.
\end{eqnarray}
It is easy to get $c_{1}(r_{m})= 0$ in the limit $r_{0}\rightarrow r_{m}$,  while $c_2(r_m)$ is complex in this limit. Furthermore, when $\beta_0= 0$, i.e., in vacuum, the $c_2(r_m)$ term in the limit $r_{0}\rightarrow r_{m}$ becomes
\begin{eqnarray}
c_{2}(r_{m})=&&\frac{96 \alpha ^2 M^3 r_m^3+\alpha  M^2 r_m^2 \left(128 \alpha ^2-12 r_m^4 \left(\sqrt{\frac{8 \alpha  M}{r_m^3}+1}-5\right)+8 \alpha  r_m^2 \left(8-11 \sqrt{\frac{8
			\alpha  M}{r_m^3}+1}\right)\right)}{\left(r_m^2 \left(\sqrt{\frac{8 \alpha  M}{r_m^3}+1}-1\right)-2 \alpha \right) \left(r_m^3+8 \alpha  M\right)^2}
\\\nonumber
&&+\frac{M \left(-6 r_m^9 \left(\sqrt{\frac{8 \alpha  M}{r_m^3}+1}-1\right)-4 \alpha  r_m^7 \left(7 \sqrt{\frac{8 \alpha  M}{r_m^3}+1}-4\right)+32 \alpha ^2
	r_m^5\right)+r_m^{10} \left(-\left(\sqrt{\frac{8 \alpha  M}{r_m^3}+1}-1\right)\right)+2 \alpha  r_m^8}{\left(r_m^2 \left(\sqrt{\frac{8 \alpha  M}{r_m^3}+1}-1\right)-2 \alpha \right) \left(r_m^3+8 \alpha  M\right)^2}.
\end{eqnarray}
Since this expression is still intricate, for the sake of clarity, let's continue to look at the form under the limit $\alpha\rightarrow 0$.
In the case $\beta_0=0$ and $\alpha=0$, $r_m=3M$, and Eq. (\ref{c2}) has a form
\begin{equation}
c_{2}(r_{m})=(6 M-r_m) r_m=9M^2,
\end{equation}
where the vacuum Schwarzschild solution is recovered.
By the discussion above, we can find that the leading term of the divergence in $f_p(z,r_{0})$ is $z^{-1}$ in the strong deflection limit, which implies $I_p(r_{0})$ diverges logarithmically.

One can separate $I_p(r_{0})$ into two parts which are the divergent
part $I_{D}(r_{0})$ and the regular part $I_{R}(r_{0})$
\begin{equation}
I_p(r_{0})=I_{D}(r_{0})+I_{R}(r_{0}).
\end{equation}
The divergent part $I_{D}(r_{0})$ is defined as
\begin{equation}
I_{D}(r_{0})\equiv \int^{1}_{0}f_{D}(z,r_{0})dz,
\end{equation}
where
\begin{equation}
f_{D}(z,r_{0})\equiv \frac{2r_{0}}{\sqrt{c_{1}(r_{0})z+c_{2}(r_{0})z^{2}}}.
\end{equation}
$I_{D}(r_{0})$ can be integrated and the result is
\begin{equation}\label{ID}
I_{D}(r_{0})=\frac{4r_{0}}{\sqrt{c_{2}(r_{0})}}\log \frac{\sqrt{c_{2}(r_{0})}+\sqrt{c_{1}(r_{0})+c_{2}(r_{0})}}{\sqrt{c_{1}(r_{0})}}.
\end{equation}
The regular part $I_{R}(r_{0})$ is defined as
\begin{equation}
I_{R}(r_{0})\equiv \int^{1}_{0}f_{R}(z,r_{0})dz,
\end{equation}
where
\begin{equation}
f_{R}(z,r_{0}) \equiv f(z,r_{0})-f_{D}(z,r_{0}).
\end{equation}

Using a similar derivation as in Ref. \cite{Tsukamoto:2016jzh}, we obtain the deflection angle $\hat{\alpha}_p(b)$ in the strong deflection limit $r_{0}\rightarrow r_{m}$ or $b \rightarrow b_{c}$ in the EGB black hole with homogeneous plasma
\begin{equation}\label{alpha1}
\hat{\alpha}_p(b)= -\bar{a}\log \left( \frac{b}{b_{c}}-1 \right) +\bar{b}+O((b-b_{c})\log (b-b_{c})).
\end{equation}
The coefficients $\bar{a}$ and $\bar{b}$ are obtained as
\begin{eqnarray}\label{abar}
&&\bar{a}=\sqrt{\frac{2B_{m}}{C_{m}\left[\frac{(CW)_{m}^{''}}{(CW)_{m}}-\frac{A_{m}^{''}}{A_{m}}\right]}},\\
&&\bar{b}=\bar{a}\log \left\{r^{2}_{m}\left[\frac{(CW)_{m}^{''}}{(CW)_{m}}-\frac{A_{m}^{''}}{A_{m}}\right]\right\} +I_{Rp}(r_{m})-\pi,\label{bbar}
\end{eqnarray}
where the subscript $m$ denotes the quantities at $r=r_m$.
In the vacuum case, i.e., $\beta_0=0$, and the coupling constant $\alpha= 0$,  $\bar{a}$ and $\bar{b}$ will reduce to the formalism in Ref. \cite{Tsukamoto:2016jzh}, which are the cofficients of a Schwarzschild black hole without plasma.

\begin{figure}
\begin{center}
		\includegraphics[width=80mm,angle=0]{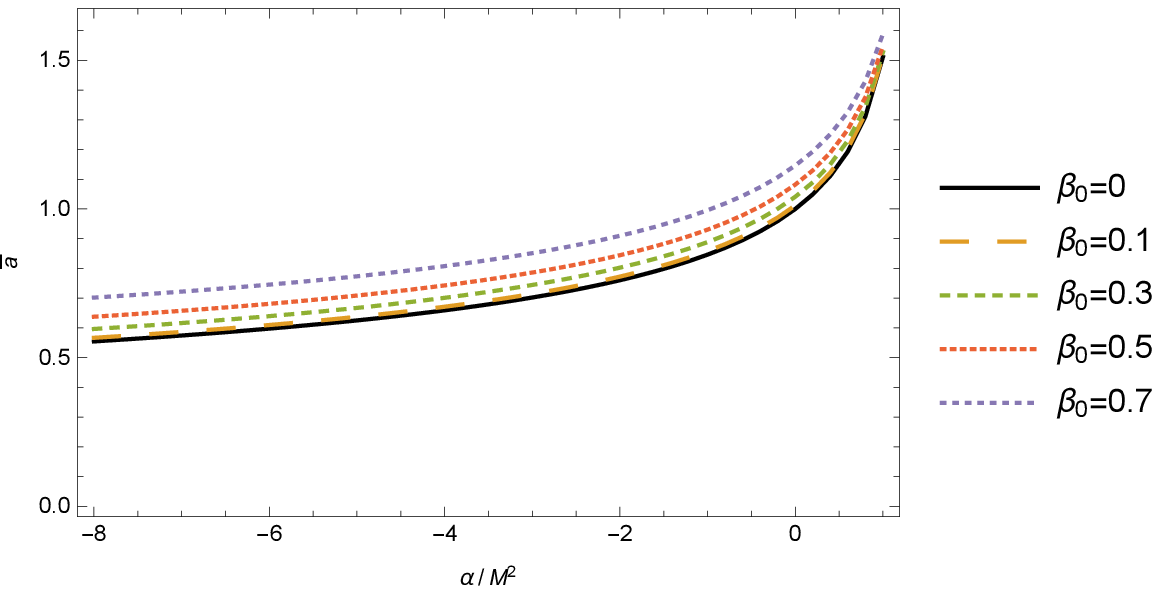}\,
		\includegraphics[width=80mm,angle=0]{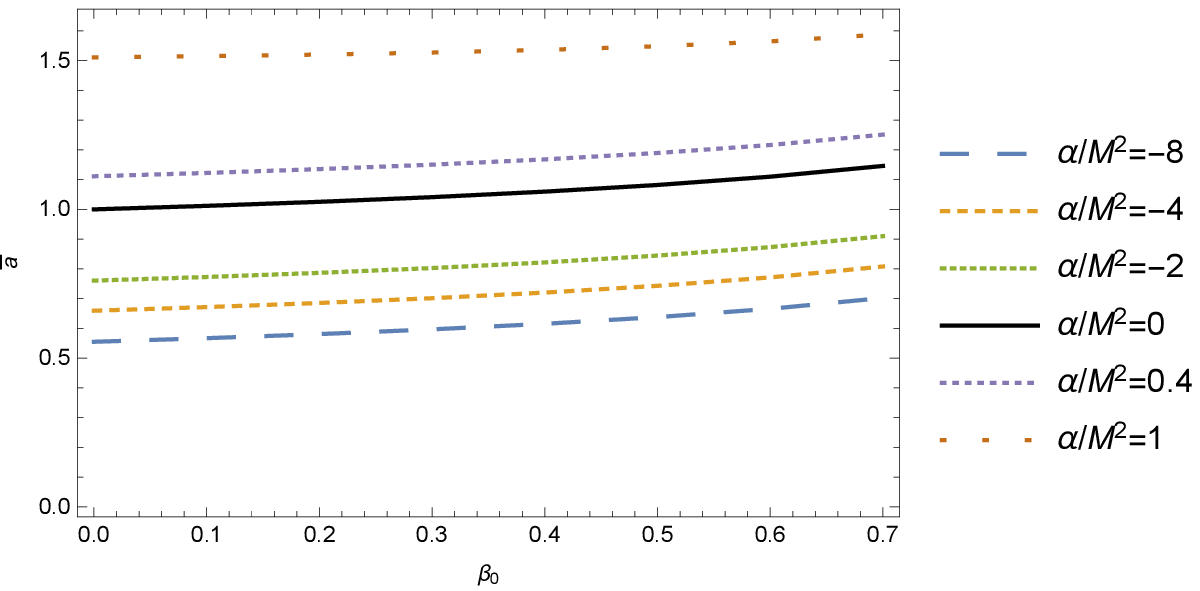}
\end{center}
\caption{Left panel: The plot of the strong field limit coefficients $\bar{a}$ as a function of $\alpha/M^2$. The five curved lines are plotted when $\beta_0=0$, $\beta_0=0.1$, $\beta_0=0.3$, $\beta_0=0.5$ and $\beta_0=0.7$ respectively.
Right panel: The plot of the strong field limit coefficients $\bar{a}$ as a function of $\beta_0$. The six curved lines are plotted when $\alpha/M^2=-8$, $\alpha/M^2=-4$, $\alpha/M^2=-2$, $\alpha/M^2=0$, $\alpha/M^2=0.4$ and $\alpha/M^2=1$ respectively.}
	\label{ab1}
\end{figure}

\begin{figure}
\begin{center}
		\includegraphics[width=80mm,angle=0]{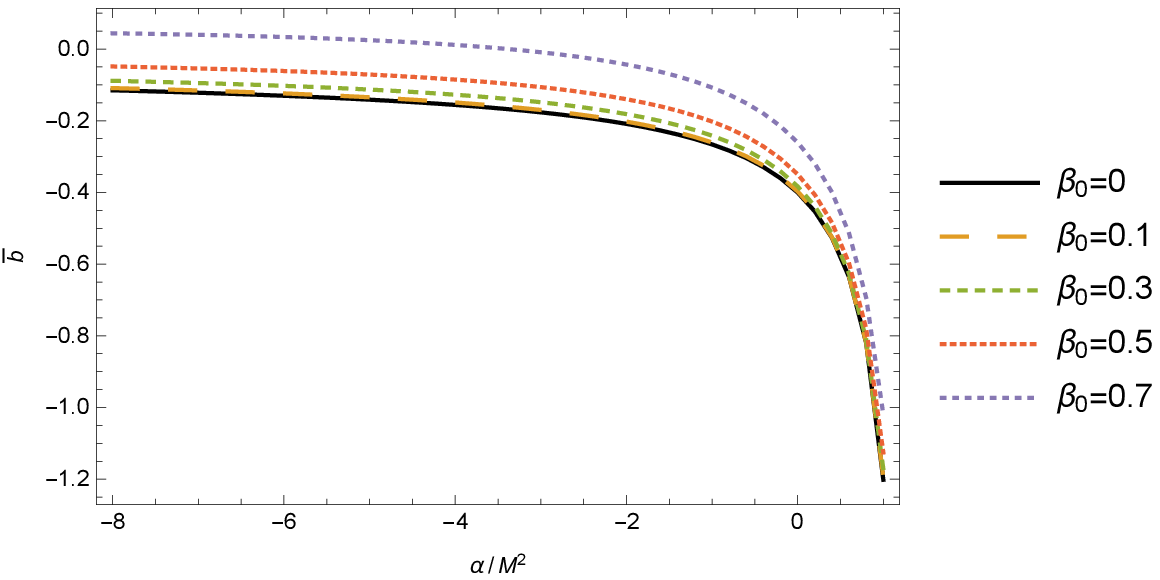}\,
		\includegraphics[width=80mm,angle=0]{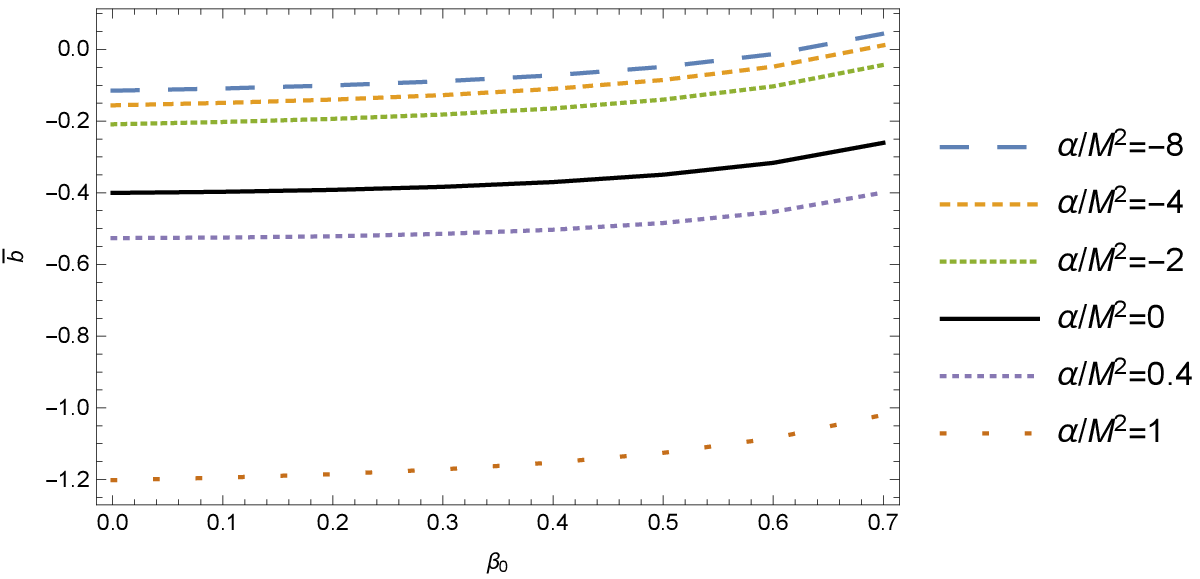}
\end{center}
\caption{Left panel: The plot of the strong field limit coefficients $\bar{b}$ as a function of $\alpha/M^2$. The five curved lines are plotted when $\beta_0=0$, $\beta_0=0.1$, $\beta_0=0.3$, $\beta_0=0.5$ and $\beta_0=0.7$ respectively.
Right panel: The plot of the strong field limit coefficients $\bar{b}$ as a function of $\beta_0$. The six curved lines are plotted when $\alpha/M^2=-8$, $\alpha/M^2=-4$, $\alpha/M^2=-2$, $\alpha/M^2=0$, $\alpha/M^2=0.4$ and $\alpha/M^2=1$ respectively.}
\label{bb1}
\end{figure}

The numerical results of the strong field limit coefficients $\bar{a}$ and $\bar{b}$ are shown in  Fig. \ref{ab1} and Fig. \ref{bb1}. 
The left panels of Fig. \ref{ab1} show the strong field limit coefficient $\bar{a}$ for $\beta_0=0$, $\beta_0=0.1$, $\beta_0=0.3$, $\beta_0=0.5$ and $\beta_0=0.7$ respectively.
We find that the strong field limit coefficient $\bar{a}$ increases with the increase of the coupling constant $\alpha/M^2$ for fixed $\beta_0$.
From the right panels of Fig. \ref{ab1}, which refers to the strong field limit coefficient $\bar{a}$ for $\alpha/M^2=-8$, $\alpha/M^2=-4$, $\alpha/M^2=-2$, $\alpha/M^2=0$, $\alpha/M^2=0.4$ and $\alpha/M^2=1$ respectively,
we find that the strong field limit coefficient $\bar{a}$ increases with the increase of the plasma parameter $\beta_0$ for fixed  $\alpha/M^2$.
In the left panels of Fig. \ref{bb1}, we illustrate the strong field limit coefficient $\bar{b}$ for $\beta_0=0$, $\beta_0=0.1$, $\beta_0=0.3$, $\beta_0=0.5$ and $\beta_0=0.7$ respectively, and show
that the strong field limit coefficient $\bar{b}$ decreases with the increase of the coupling parameter $\alpha/M^2$ for fixed $\beta_0$.
From the right panels of Fig. \ref{bb1}, which refers to the strong field limit coefficient $\bar{b}$ for $\alpha/M^2=-8$, $\alpha/M^2=-4$, $\alpha/M^2=-2$, $\alpha/M^2=0$, $\alpha/M^2=0.4$ and $\alpha/M^2=1$ respectively, we find that the strong field limit coefficient $\bar{b}$ increases with the increase of the plasma parameter $\beta_0$ for fixed  $\alpha/M^2$.
Obviously, the strong field limit coefficients $\bar{a}$ and $\bar{b}$ are influenced by the choice of coupling constant $\alpha$ and plasma parameter $\beta_0$.

%%%%%%%%%%%%%%%%%%%%%%%%%%%

\section{Observables in the strong deflection limit}

In this section we consider the observables of the strong gravitational lensing by the EGB black hole in the presence of a uniform plasma.
We are interested in the case where the observer, the lens and the source are nearly in alignment, and the source and the observer are very far from the lens. 
The distance between the lens and the source, and the distance between the lens and the observer are represented by $D_{LS}$ and $D_{OL}$ respectively. 
$D_{OS}$ is the distance between the observer and the source, and $D_{OS}=D_{LS}+D_{OL}$.
$\beta$ denotes the angular position with respect to the optical axis of the source.
$\theta$ is the angular position with respect to the optical axis of the image and can be expressed as $\theta=b/D_{OL}$. 
Thus the lens eqaution can be written as \cite{Bozza:2002zj}
\begin{equation}\label{beta}
\beta=\theta-\frac{D_{LS}}{D_{OS}}\triangle\alpha_{n},
\end{equation}
where $\triangle\alpha_{n}=\alpha-2n\pi$ is the offset of deflection angle, and $n$ denotes the loop numbers of the light ray around the light sphere.

The angular position $\theta_{n}$ between the lens and the $n$-th relativistic image and the magnification of the $n$-th relativistic image $\mu_{n}$ can be obtained approximately as
\begin{eqnarray}\label{theta}
&&\theta_{n}=\theta^{0}_{n}+\frac{b_{c}(\beta-\theta_{n}^{0})D_{OS}}{\bar{a}D_{LS}D_{OL}}\exp\left({\frac{\bar{b}-2n\pi}{\bar{a}}}\right),\\
&&\mu_{n}=\frac{b_{c}^{2}D_{OS}}{\bar{a}\beta D^{2}_{OL}D_{LS}}
\exp\left({\frac{\bar{b}-2n\pi}{\bar{a}}}\right)\left[1+\exp\left({\frac{\bar{b}-2n\pi}{\bar{a}}}\right)\right].\label{mun}
\end{eqnarray}
Here $\theta_{n}^{0}$ is the angular position corresponding to the case that the light ray winds completely $2n\pi$ and can be expressed as 
\begin{equation}\label{theta1}
\theta_{n}^{0}=\frac{b_{c}}{D_{OL}}\left[1+\exp\left({\frac{\bar{b}-2n\pi}{\bar{a}}}\right)\right].
\end{equation}
In the limit $n\rightarrow\infty$, we can find the relation between 
the critical impact parameter $b_{c}$ and the asymptotic position $\theta_{\infty}$  approached by a set of images
\begin{equation}\label{theta2}
\theta_{\infty}=\frac{b_{c}}{D_{OL}}.
\end{equation}
Since the outermost relativistic image is the brightest, one can use the observable $s$ to describe the separation between this first image $\theta_{1}$ and all the others packed images at $\theta_{\infty}$ \cite{Bozza:2002zj}. The other observable $\mathcal{R}$ represents the ratio of the received flux beteween this first image and all the others images \cite{Bozza:2002zj}. 
Using Eqs. (\ref{theta}) and (\ref{mun}), the angular separation $s$ and the ratio of the flux $\mathcal{R}$ can be obtained as
\begin{eqnarray}\label{sr}
&&s=\theta_{1}-\theta_{\infty}=\theta_{\infty} \exp\left({\frac{\bar{b}-2\pi}{\bar{a}}}\right),\\
&&\mathcal{R}=\frac{\mu_{1}}{\sum\limits_{n=2}^{\infty}\mu_{n}}=\exp\left({\frac{2\pi}{\bar{a}}}\right).
\end{eqnarray}
If the observables  $s$, $\theta_{\infty}$ and $\mathcal{R}$ are available, the coefficients $\bar{a}$ and $\bar{b}$ in the strong deflction limit and the critical impact parameter $b_{c}$ can be obtained easily by
\begin{eqnarray}\label{ab2}
&& \bar{a}=\frac{2\pi}{\log \mathcal{R}},\\
&& \bar{b}=\bar{a}\log(\frac{\mathcal{R}s}{\theta_{\infty}}),\\
&& b_{c}=\theta_{\infty}D_{OL}.
\end{eqnarray}
Then one can numerically compute the above value by measuring  the observables  $s$, $\theta_{\infty}$ and $\mathcal{R}$  and study their difference with the corresponding theoretical coefficients.

Let's take the supermassive black hole M87$^{*}$ as an example.  The results from the EHT show that the angular diameter of the shadow of M87$^{*}$ is $(42\pm 3) \mu$as, and the observed shadow is almost circular which is supported by the fact that the axis ratio is smaller than $4/3$ and the deviation from circularity is less than 10\% \cite{Akiyama:2019cqa, Akiyama:2019fyp, Akiyama:2019eap}.
Therefore, the image of M87$^{*}$ is nearly circular due to the relatively small value of spin and low inclination angle $\sim 17 ^\circ$ of the source \cite{Walker:2018muw}. So it is reasonable to choose spherically symmetric metric as  an approximation to discuss the strong gravitational lensing of M87$^{*}$.
Meanwhile, the mass of M87$^{*}$ is estimated by the EHT collaboration as  $M=(6.5\pm 0.7)\times 10^{9}~M_{\odot}$ \cite{Akiyama:2019eap}. Note that the mass of M87$^{*}$ is also estimated to be $6.2^{+1.1}_{-0.5}\times10^9$ $M_{\odot}$  and $3.5^{+0.9}_{-0.3}\times10^9$ $M_{\odot}$   by the stellar dynamics \cite{Gebhardt:2011yw} and gas dynamics measurements \cite{Walsh:2013uua}, respectively.
In addition, the distance $D_{OL}$ of M87$^{*}$ from us is estimated to be $D_{OL}$=(16.8$\pm$0.8) Mpc from stellar population measurements \cite{Blakeslee:2009tc, Bird:2010rd, Cantiello:2018ffy}.

\begin{figure}
\begin{center}
		\includegraphics[width=80mm,angle=0]{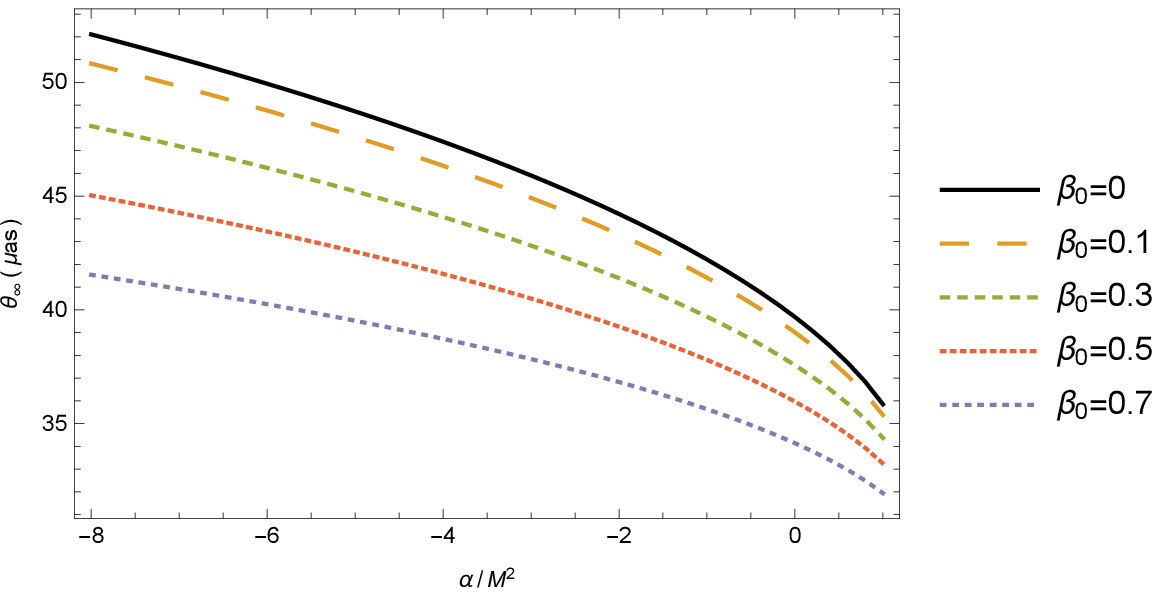}\,
		\includegraphics[width=80mm,angle=0]{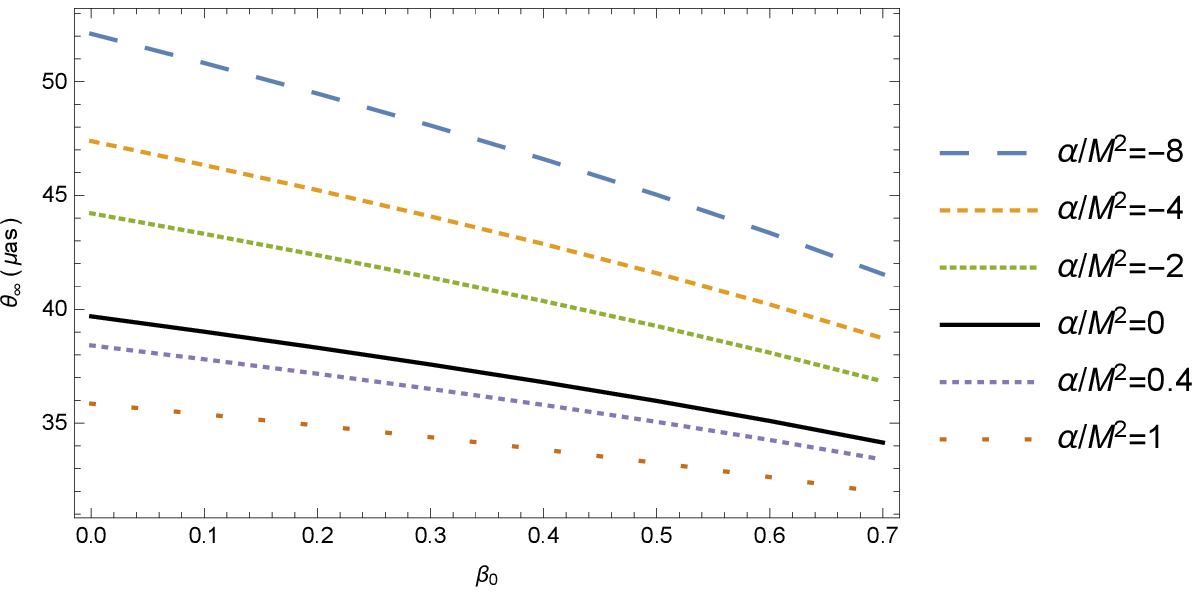}
\end{center}
\caption{Left panel: The plot of the angular image position $\theta_{\infty}$ as a function of $\alpha/M^2$. The five curved lines are plotted when $\beta_0=0$, $\beta_0=0.1$, $\beta_0=0.3$, $\beta_0=0.5$ and $\beta_0=0.7$ respectively.
Right panel: The plot of the angular image position $\theta_{\infty}$ as a function of $\beta_0$. The six curved lines are plotted when $\alpha/M^2=-8$, $\alpha/M^2=-4$, $\alpha/M^2=-2$, $\alpha/M^2=0$, $\alpha/M^2=0.4$ and $\alpha/M^2=1$ respectively.}
\label{th}
\end{figure}

\begin{figure}
\begin{center}
		\includegraphics[width=80mm,angle=0]{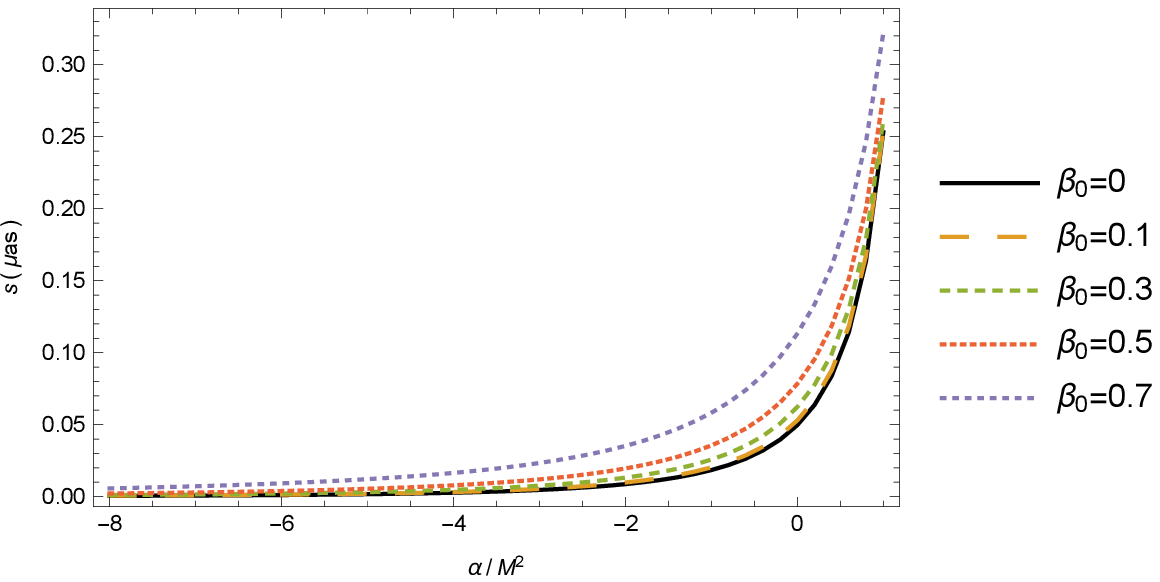}\,
		\includegraphics[width=80mm,angle=0]{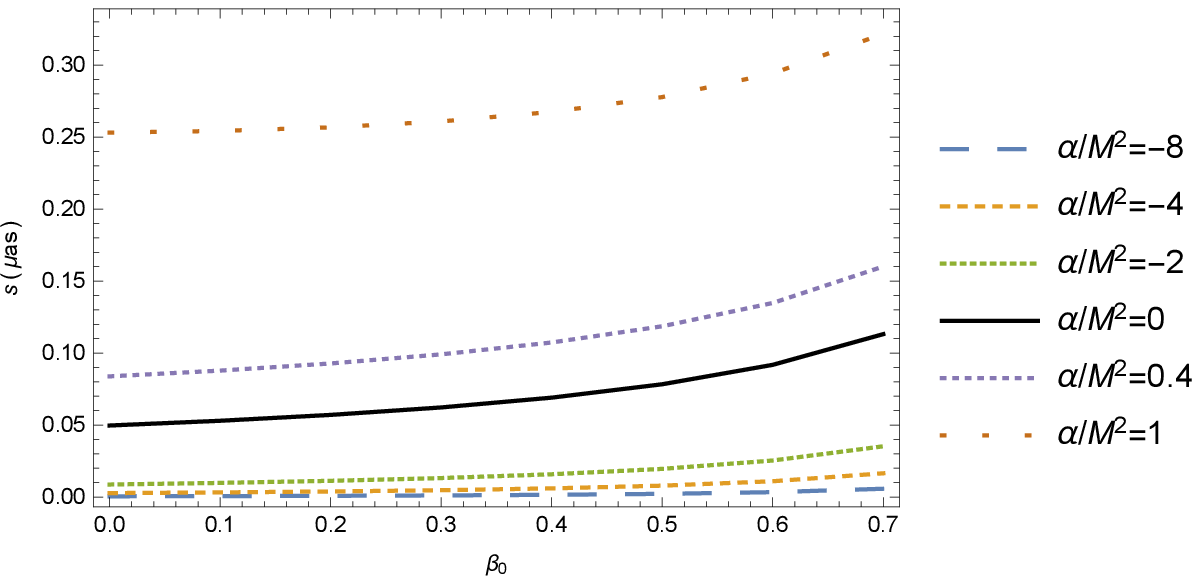}
\end{center}
\caption{Left panel: The plot of the angular image separation $s$ as a function of $\alpha/M^2$. The five curved lines are plotted when $\beta_0=0$, $\beta_0=0.1$, $\beta_0=0.3$, $\beta_0=0.5$ and $\beta_0=0.7$ respectively.
Right panel: The plot of the angular image separation $s$ as a function of $\beta_0$. The six curved lines are plotted when $\alpha/M^2=-8$, $\alpha/M^2=-4$, $\alpha/M^2=-2$, $\alpha/M^2=0$, $\alpha/M^2=0.4$ and $\alpha/M^2=1$ respectively.}
\label{s}
\end{figure}

\begin{figure}
\begin{center}
		\includegraphics[width=80mm,angle=0]{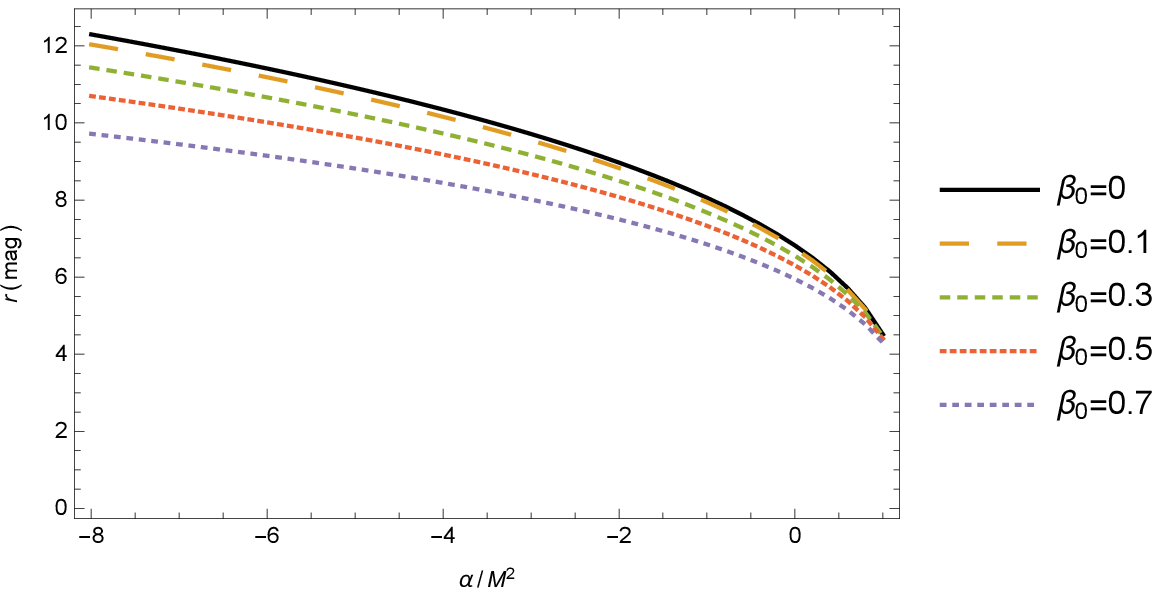}\,
		\includegraphics[width=80mm,angle=0]{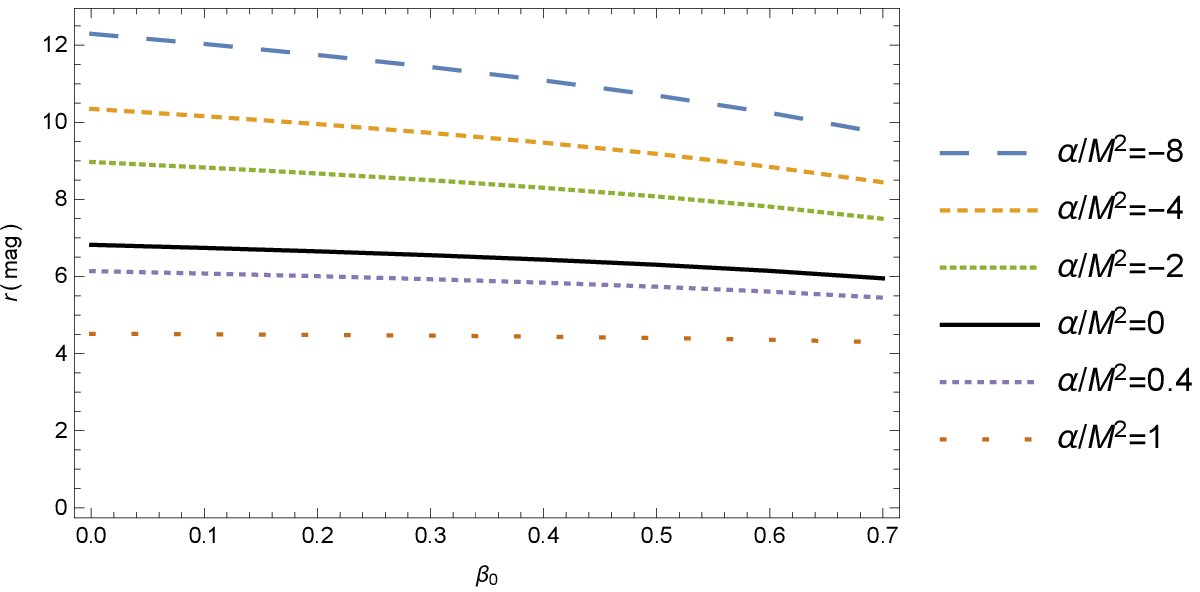}
\end{center}
\caption{Left panel: The plot of the relative magnifications $r$ as a function of $\alpha/M^2$. The five curved lines are plotted when $\beta_0=0$, $\beta_0=0.1$, $\beta_0=0.3$, $\beta_0=0.5$ and $\beta_0=0.7$ respectively.
Right panel: The plot of the relative magnifications $r$ as a function of $\beta_0$. The six curved lines are plotted when $\alpha/M^2=-8$, $\alpha/M^2=-4$, $\alpha/M^2=-2$, $\alpha/M^2=0$, $\alpha/M^2=0.4$ and $\alpha/M^2=1$ respectively.}
\label{dm}
\end{figure}

In the following the lens is supposed to be M87$^{*}$ which is described by the EGB black hole. For simplicity, we use the following data from M87$^{*}$, $D_{OL}$=16.8 Mpc and $M=$6.5$\times10^9$ $M_{\odot}$. With these data we can estimate the values of the  angular image position $\theta_{\infty}$, the angular image separation $s$ and the relative magnifications $r$ of the relativistic images which is defined as $r=2.5\log_{10}\mathcal{R}$.
Figs. \ref{th}-\ref{dm} show the behaviors of these observables and the 
influences on them by the choince of coupling constant $\alpha$ and plasma parameter $\beta_0$.
The left panels of Fig. \ref{th} show the value of $\theta_{\infty}$ as function of
$\alpha/M^2$ for $\beta_0=0$, $\beta_0=0.1$, $\beta_0=0.3$, $\beta_0=0.5$ and $\beta_0=0.7$ respectively. As the coupling constant $\alpha/M^2$ increases,
the angular image position decreases for fixed $\beta_0$. 
The right panels of Fig. \ref{th} show the value of $\theta_{\infty}$ as function of
$\beta_0$ for $\alpha/M^2=-2$, $\alpha/M^2=0$, $\alpha/M^2=0.4$ and $\alpha/M^2=1$ respectively. As the plasma parameter $\beta_0$ increases,
the angular image position decreases for fixed $\alpha/M^2$.
 
As shown in the left panels of Fig. \ref{s}, the value of $s$ is expressed as a function of $\alpha/M^2$ for $\beta_0=0$, $\beta_0=0.1$, $\beta_0=0.3$, $\beta_0=0.5$ and $\beta_0=0.7$ respectively. As the coupling constant $\alpha/M^2$ increases, the angular image separation increases for fixed $\beta_0$.
It is shown that in the right panels of Fig. \ref{s}, the value of $s$ is expressed as function of $\beta_0$ for $\alpha/M^2=-2$, $\alpha/M^2=0$, $\alpha/M^2=0.4$ and $\alpha/M^2=1$ respectively. As the plasma parameter $\beta_0$ increases, the angular image separation increases for fixed $\alpha/M^2$.
Interestingly, as $\alpha/M^2\rightarrow-8$ for different plasma parameter $\beta_0$, the angular image separation converges to the value at $\alpha/M^2=-8$ in the vacuum case,
which means plasma has little effect on the angular image separation. 

In the left panels of Fig. \ref{dm}, we show the value of relative magnifications as function of
$\alpha/M^2$ for $\beta_0=0$, $\beta_0=0.1$, $\beta_0=0.3$, $\beta_0=0.5$ and $\beta_0=0.7$ respectively. As the coupling constant $\alpha/M^2$ increases,
the relative magnifications  decreases for fixed $\beta_0$. 
In the right panels of Fig. \ref{dm}, we show the value of relative magnifications as function of
$\beta_0$ for $\alpha/M^2=-2$, $\alpha/M^2=0$, $\alpha/M^2=0.4$ and $\alpha/M^2=1$ respectively. As the plasma parameter $\beta_0$ increases,
the relative magnifications decreases for fixed $\alpha/M^2$.
We find that as $\alpha/M^2\rightarrow 1$ for different plasma parameter $\beta_0$, the relative magnifications converges to the value at $\alpha/M^2=1$ in the vacuum case,
which means plasma has little influence on the relative magnifications.

In Table \ref{tab1}, we list the numerical estimates of the observables as well as strong field limit coefficients of a EGB black hole in uniform plasma. 
The parameter $\beta_0=0$  corresponds to the case of the EGB black hole in vacuum and $\alpha=0$ means the case of Schwarzschild
black hole in homogeneous plasma.
From Table \ref{tab1}, we can easily obtain the differences between the Schwarzschild black hole and the EGB  black hole, as well as the EGB  black hole with various plasma parameter.


%%%%%%%%%%%%%%%%%%%%%%%%%%%%%%%%%%%%%%%%%%%%%%%%%%%%%%%%%%%%%%%%%%%%%%%%%%%%
\begin{table}[h]
\begin{center}
\begin{tabular}{cccccccc}
\hline\hline
$\beta_0$ & $\alpha/M^2$ & $\theta_{\infty}$($\mu\text{as}$) & $s$($\mu \text{as}$) & $r(\text{mag})$  & $b_{c}/R_{s}$ & $\bar{a}$ & $\bar{b}$ \\
\hline
 & -8 & 52.10 & 0.000511 & 12.30 & 3.41 & 0.555 & -0.1151 \\
 & -4 & 47.39 & 0.00272  & 10.35 & 3.10 & 0.659 & -0.1558 \\
 & -2 & 44.21 & 0.00867  & 8.97  & 2.89 & 0.760 & -0.2087 \\
0& 0  & 39.69 & 0.0497   & 6.82  & 2.60 & 1     & -0.4003 \\
 & 0.4& 38.41 & 0.0838   & 6.14  & 2.51 & 1.111 & -0.5264 \\
 & 1  & 35.85 & 0.253    & 4.51  & 2.35 & 1.511 & -1.2017 \\\hline
   & -8 & 50.82 & 0.000644 & 12.03 & 3.33 & 0.567 & -0.1088 \\
   & -4 & 46.33 & 0.00320  & 10.16 & 3.03 & 0.671 & -0.1490 \\
   & -2 & 43.31 & 0.00980  & 8.83  & 2.83 & 0.773 & -0.2024 \\
0.1& 0  & 39.01 & 0.0530   & 6.74  & 2.55 & 1.012 & -0.3971 \\
   & 0.4& 37.80 & 0.0878   & 6.08  & 2.47 & 1.123 & -0.5249 \\
   & 1  & 35.38 & 0.254    & 4.50  & 2.32 & 1.515 & -1.1947 \\\hline
   & -8 & 48.08 & 0.000111 & 11.43 & 3.15 & 0.597 & -0.08858 \\
   & -4 & 44.08 & 0.00473  & 9.73  & 2.89 & 0.701 & -0.1274 \\
   & -2 & 41.39 & 0.00132  & 8.50  & 2.71 & 0.803 & -0.1815 \\
0.3& 0  & 37.57 & 0.0622   & 6.55  & 2.46 & 1.041 & -0.3833 \\
   & 0.4& 36.50 & 0.0991   & 5.93  & 2.39 & 1.150 & -0.5144 \\
   & 1  & 34.38 & 0.261    & 4.47  & 2.25 & 1.527 & -1.1715 \\\hline   
   & -8 & 45.03 & 0.00220  & 10.70 & 2.95 & 0.638 & -0.04854 \\
   & -4 & 41.58 & 0.00788  & 9.18  & 2.72 & 0.743 & -0.08515 \\
   & -2 & 39.26 & 0.00196  & 8.08  & 2.57 & 0.845 & -0.1400 \\
0.5& 0  & 35.97 & 0.0783   & 6.30  & 2.35 & 1.082 & -0.3494 \\
   & 0.4& 35.05 & 0.119    & 5.73  & 2.29 & 1.190 & -0.4843 \\
   & 1  & 33.25 & 0.278    & 4.41  & 2.18 & 1.548 & -1.1255 \\\hline   
   & -8 & 41.54 & 0.00574  & 9.72  & 2.72 & 0.702 &  0.04408 \\
   & -4 & 38.73 & 0.0165   & 8.44  & 2.54 & 0.808 &  0.01158 \\
   & -2 & 36.83 & 0.0352   & 7.50  & 2.41 & 0.910 & -0.04313 \\
0.7& 0  & 34.14 & 0.113    & 5.95  & 2.24 & 1.146 & -0.2606 \\
   & 0.4& 33.40 & 0.160    & 5.45  & 2.19 & 1.251 & -0.3993 \\
   & 1  & 31.95 & 0.322    & 4.30  & 2.09 & 1.588 & -1.0194 \\\hline
\hline
\end{tabular}
\caption{Numerical estimation for the observables and the strong deflection limit coefficients for EGB black holes supposed to describe the supermassive black hole M87$^{*}$. $R_S=2GM/c^2$ is the Schwarzschild radius.}\label{tab1}
	\end{center}
\end{table}
%%%%%%%%%%%%%%%%%%%%%%%%%%%%%%%%%%%%%%%%%%%%%%%%%%%%%%%%%%%%%%%%%%%%%%%%%%%%%%

\section{Conclusions}

In this work, we have investigated the strong gravitational lensing generated by a 4-dimensional Einstein-Gauss-Bonnet black hole in a plasma. 
In the presence of plasma around the black hole, the trajectory of a photon differs from the null geodesic in vacuum, resulting in the changes of the deflection angle of light.
Using Hamilton's equation of the light ray in plasma with a frequency dependent refraction index, we have derived the equation of motion for light rays in the novel $4$-dimensional EGB black hole. 
Furthermore, we numerically obtained the theoretical strong field limit parameters for the lensing by the black hole in a uniform plasma. Among these parameters we found that the radius of the photon sphere $r_{m}$,
the critical impact parameter $b_c$ and the strong field limit coefficient $\bar{b}$ 
decrease monotonically, while the strong field limit coefficient $\bar{a}$ 
increases, with the increase of the coupling constant $\alpha/M^2$ for fixed value of plasma parameter $\beta_0$.
On the other hand, for a fixed value of the coupling constant $\alpha/M^2$,   with the increase of the plasma parameter $\beta_0$, $r_{m}$, $\bar{a}$ and  $\bar{b}$ increase, but  $b_c$ decreases monotonically.
Modelling the supermassive $\mathrm{M}87^{\ast}$ with this EGB black hole,
we have estimated the observables including the angular image position $\theta_{\infty}$, the angular image separation $s$ and the relative magnifications $r$ of the relativistic images in the uniform plasma.
We have shown that among these observables, when the coupling parameter $\alpha/M^2$ increases for fixed plasma parameter $\beta_0$, the angular image position $\theta_{\infty}$ and the relative magnifications $r$ decrease,
while the angular image separation $s$ increases. 
When the  plasma parameter $\beta_0$ increases for fixed coupling constant $\alpha/M^2$, the angular image position $\theta_{\infty}$ and the relative magnifications $r$ decrease, but the angular image separation $s$ increases.
Above all, both the coupling constant $\alpha$ and plasma parameter $\beta_0$ have significant effects on the parameters and observables in strong gravitational lensing.
Interestingly, it is found that  plasma has little effect on the angular image separation as $\alpha/M^2\to -8$ and  the relative magnifications as $\alpha/M^2\to 1$, respectively. 

Theoretically we can use the observations on strong gravitational lensing to test this modified gravity, although the relativistic images of strong gravitational lensing are so faint that it is hard to detect. However, with the improvement of technology, we wish observations in the future may provide the opportunity to distinguish the EGB black hole from those in general relativity. Finally, it is worth noting that in this paper we ignore the rotation of M87$^{*}$ and use the spherically symmetric metric to  provide some hints about its gravitational lensing signals. The gravitational lensing effect of rotating EGB black hole deserves a new work in the future.

\begin{acknowledgments}
	{{This work is supported in part by Science and Technology Commission of Shanghai Municipality under Grant No. 12ZR1421700 and Shanghai Normal University  KF201813.}}
\end{acknowledgments}

\textbf{Note added:} After this work is completed, we are  aware of a similar work by 
Islam et al. \cite{Islam:2020xmy}, which appeared in arXiv a couple of days before.
They focus on the strong gravitational lensing in vacuum, in which the variation range of the coupling constant is $0\le {\alpha}/{M^2}\le 0.019$, while we discuss the strong gravitational lensing in homogeneous plasma, in which the variation range of coupling constant is $-8\le {\alpha}/{M^2}\le 1$.
Our results agree with theirs where we overlap.

\end{document}